# Signatures of magnetism in zigzag graphene nanoribbon embedded in *h*-BN lattice

Chengxin Jiang,[1,#] Hui Shan Wang,[1,2,3,#*] Chen Chen,[1,#] Lingxiu Chen,[4] Xiujun Wang,[1,2] Yibo Wang,[1] Ziqiang Kong,[1,2] Yuhan Feng,[1,2] Yixin Liu,[1] Yu Feng,[1,2] Chenxi Liu,[1,2] Yu Zhang,[1,2] Zhipeng Wei,[5] Maosen Guo,[6] Aomei Tong,[6] Gang Mu,[1,2] Yumeng Yang,[7] Kenji Watanabe,[8] Takashi Taniguchi,[9] Wangzhou Shi,[3] Haomin Wang[1,2*]

[1] State Key Laboratory of Functional Materials for Informatics, Shanghai Institute of Microsystem and Information Technology, Chinese Academy of Sciences, Shanghai 200050, China.
[2] Center of Materials Science and Optoelectronics Engineering, University of Chinese Academy of Sciences, Beijing 100049, China.
[3] Key Laboratory of Optoelectronic Material and Device, Department of Physics, Shanghai Normal University, Shanghai 200234, China.
[4] School of Materials Science and Physics, China University of Mining and Technology, Xuzhou 221116, China.
[5] State Key Laboratory of High Power Semiconductor Lasers, Changchun University of Science and Technology, Changchun, 130022, People's Republic of China.
[6] CIQTEK Co,. Ltd., Hefei 230000, China.
[7] School of Information Science and Technology, ShanghaiTech University, Shanghai 201210, China.
[8] Research Center for Functional Materials, National Institute for Materials Science, 1-1 Namiki, Tsukuba, 305-0044 Japan.
[9] International Center for Materials Nanoarchitectonics, National Institute for Materials Science, 1-1 Namiki, Tsukuba 305-0044, Japan.

E-mail: hswang2024@shnu.edu.cn; hmwang@mail.sim.ac.cn

**Abstract:** Zigzag edges of graphene have long been predicted to exhibit magnetic electronic state near the Fermi level, which can cause spin-related phenomena and offer unique potentials for graphene-based spintronics. However, the magnetic conduction channels along these edges have yet been reported experimentally. Here, we report the observation on signatures of magnetism in zigzag graphene nanoribbons (zGNRs) embedded in hexagonal boron nitride (*h*-BN). The in-plane bonding with BN can stabilize the edges of zGNRs, and thus enable a direct probing of the intrinsic magnetism. Firstly, the presence of magnetism of a zGNR was confirmed by scanning NV center microscopy. And then, zGNR was fabricated into a transistor with a width of ~9 nm wide and a channel length of sub-50 nm. By performing magneto-transport measurements, Fabry-Pérot interference patterns were observed in the transistor at 4 Kelvin, which indicates a coherent transport through the channel. A large magnetoresistance of ~175 Ω, corresponding to a ratio of ~1.3 %, was observed at the same temperature. More importantly, such magneto-transport signal is highly anisotropic on the magnetic field direction, and its appearance extends well above room temperature. All these evidences corroborate the existence of robust magnetic ordering in the edge state of zGNR. The findings on zGNR embedded in *h*-BN provide an effective platform for the future exploration of graphene-based spintronic devices.

Zigzag graphene nanoribbons (zGNRs) are known as quasi one-dimensional strips of graphene associated by parallel zigzag edges. These localized zigzag edges are predicted to possess a unique magnetic state near the Fermi level[1,2] that can be promising for spintronic device applications. Recent progresses in GNRs synthesis enabled the fabrication of atomically precise GNRs, allowing the electrical manipulation and detection of edge states. Symmetry protected topological phases,[3-5] zero-mode metallicity[6] and spin-polarized edge states[7] are subsequently revealed via advanced scanning probe spectroscopy in GNRs. However, a direct experimental probing of the magnetism by electrical methods in zigzag edge structures still remains elusive. Limited by surface diffusivity and kinetic factors on catalytic substrates, GNRs[8,9] made from bottom-up assembly have length typically in the range of tens of nanometers, which is too short to be technologically practical for reliable device fabrication. Moreover, the open edges of zGNRs[7,10-14] are of relatively high chemical reactivity, and can thus cause instability or inhomogeneous doping in electronic properties.[16-18] Finally, the edge states in narrow zGNRs could become antiferromagnetically coupled so strongly that imposes difficulties detect sizable electrical signal under a moderate magnetic field.[8] All the above reasons hurdle the direct transport measurement on the magnetism in GNR electronic device.

Successful fabrication of oriented zGNRs embedded in hexagonal boron nitride (*h*-BN) could potentially overcome the aforementioned difficulties and enable exploration of spin transport in the ultimate limit of miniaturization.[19] The zigzag-oriented *h*-BN trenches were produced by zinc nanoparticles cutting.[20] Subsequently, the zigzag-oriented *h*-BN trenches were filled with zGNR by gaseous catalyzed chemical vapor deposition.[21] These zGNRs are now stabilized by in-plane graphene–BN bonding,[22,23] having a C-B interface at one edge and a C-N interface at the opposite edge. And thus narrow zGNRs were predicted to exhibit an insulating antiferromagnetic state[24-28] with negligible contribution to electrical signal, similar to hydrogen terminated zGNRs[2]. On another aspect, the asymmetric structure of zGNR/BN interface could in turn weaken the super-exchange interaction between the two inequivalent carbon edge states, so that now only a moderate magnetic field is required to manipulate the magnetic order of zGNR.[28] More interestingly, it was further predicted that the edges of zGNRs could become a half-metal once an in-plane homogeneous electric field is applied across the width direction.[7] Unfortunately, the transverse electric field required is experimentally too high (normally higher than one Volt per nm) to realize half-semimetallicity in ultra-narrow zGNRs.[7,29] However, in a slightly wider zGNR, the super-exchange interaction between edge states could become very weak. In addition, a built-in electric field is naturally created by the difference of electrostatic potentials at the C-B and C-N interfaces, which could promote the formation of half-semimetallicity in zGNRs' edge states.[30] As such, the zGNRs embedded in *h*-BN provide a unique material platform to detect the intrinsic edge state magnetism by transport measurement in a proper magnetic field, which could largely deepen the understanding of zGNR's magnetic properties.

A recently emerged scanning NV center microscope (SNVM) has become a powerful tool under ambient conditions to study spin textures in various magnetic materials.[31-34] By measuring the local magnetic field quantitatively via optical detection magnetic resonance (ODMR)[35,36], the SNVM acquire nanoscale spatial resolution of the sample's stray fields in a magnetic sensitivity of $\mu T/\sqrt{\mathrm{Hz}}$. [37,38]

Firstly, we examine the presence of magnetism in zGNRs using SNVM, and then the results are illustrated in figure 1. Figure 1b shows an AFM image of the region with zGNRs. Figure 1d shows the stray field in ISO-B mode for the same region shown in figure 1b. The contours of the magnetic domains extracted from PL mapping clearly shows the obvious contrast at the position of the zGNR. To exclude the contribution of other origins to the PL signal, full-B mode imaging is adapted.[33] As shown in figure 1e, the full-B mode scanning exhibits a clearer and more distinct magnetic field distribution. Crucially, its contours are consistent with those of the zGNR measured by AFM. The results of both scanning modes of SNVM confirm the magnetism of zGNR embedded in *h*-BN.

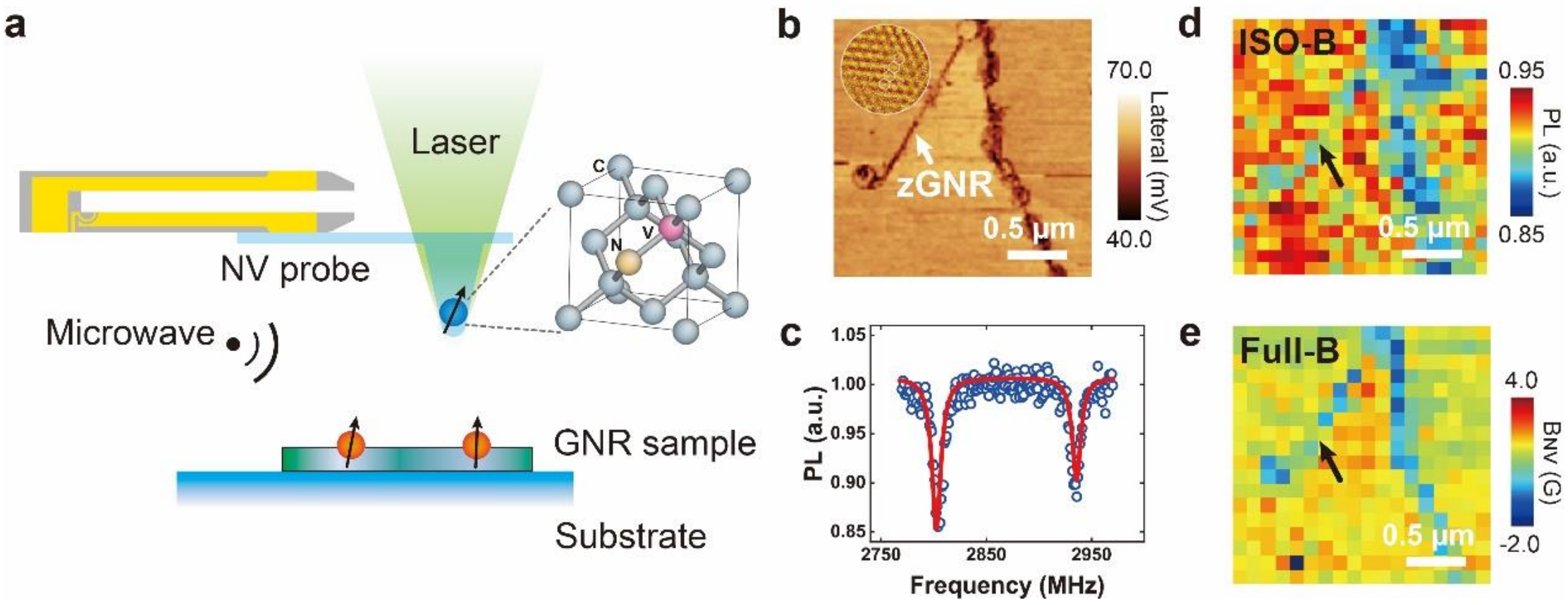

**Figure 1. Probing the magnetism of zGNRs embedded in *h*-BN with a scanning nitrogen-vacancy (NV) microscope (SNVM) under ambient conditions.** (a) The experimental setup of the SNVM. The single NV defect locates ~6 nm away from the top of diamond tip as shown in the inset; (b) A lateral force image of the zGNR acquired by AFM. The inset shows the lattice resolution friction image of *h*-BN which indicates the orientation of GNR; (c) Optically detected magnetic resonance (ODMR) spectrum of NV center at an external vertical magnetic field (~40 G). (d) The result of iso-B mode scanning at a fixed microwave frequency of 2801 MHz. Under the iso-B mode, signal of the magnetic field along the NV axis is collected during the scanning. (e) The full-B mode scanning result of zGNR, which exhibits the quantitative stray field magnitude along the NV center. (b-e) are obtained in the same region.

Embedded zGNRs were then transferred onto Si/$SiO_2$ substrate together with underlying *h*-BN for device fabrication. SEM image of an embedded zGNR is given in **Fig. S1**. The discretely distributed GNRs allow the fabrication of transistors from the individual zGNR. Figure 2a shows the schematics of a typical two-terminal zGNR device. One embedded zGNR is used as the channel material, and the channel length ($L$) is defined by the distance between metallic electrodes. The right panel in figure 1b shows the sketch of a zGNR embedded in the top layer of *h*-BN substrate. Figure 1b-d shows the electrical properties of a ~9 nm-wide zGNR field effect transistor (FET) with $L$ of ~50 nm at 4 K. As can be seen, the zGNR is of metallic nature with a conductance oscillating in the whole range of $V_{gate}$ swept. It is noticed that the conductance $G$ is about ~2.64 $e^2/h$ at $V_{gate}$ = -60 V. It indicates that the contact resistance is greatly reduced and charge carriers keep coherence in their transport through the zGNR channel. It is found that there is a conductance peak near $V_{gate}$ = 10 V, which may be due to the presence of electronic states at the opposite edges of zGNR. As indicated by black arrows, obvious Fabry-Pérot (FP) like interference patterns of differential conductance are observed when $V_{gate}$ is smaller than -35V (see figure 2c) and in the range from 25 to 35 V (see figures 2d). One extracts a value of ~30 meV for the level spacing between the neighboring resonance peaks along $V_{sd}$. The value can be obtained by setting the round trip phase shift $2LeV_{sd}/\hbar\upsilon_F$ equal to $2\pi$, which is in good agreement with the shadow-mask defined spacing of ~50 nm between the Pd electrodes for a Fermi velocity of $7.25\times10^5$ m/s. The diamonds are not completely symmetric, which could be explained through slightly asymmetrical contacts. The appearance of FP like interference pattern, together with a high conductance, indicates that the transport of charge carriers is almost ballistic in the as-fabricated zGNR FET. This ballistic transport nature of FET lays a solid foundation for probing the intrinsic magnetism of the edge states in zGNR.

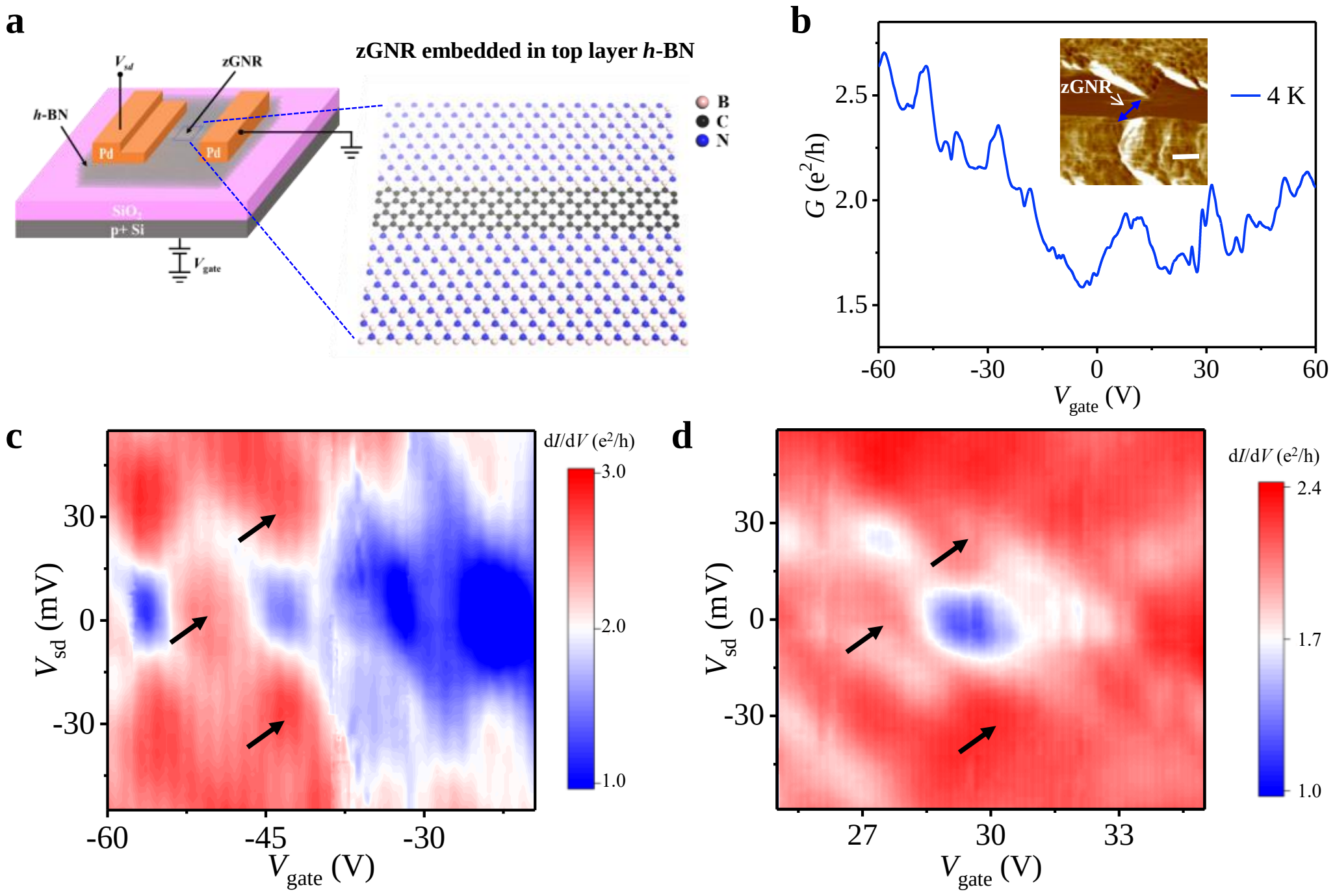

**Figure 2. A ~9 nm-wide zGNR embedded in *h*-BN and its electrical properties.** (a) Schematics of a zGNR device sitting on $SiO_2$/Si substrate where highly doped Si substrate acts as a global back gate.(b) Conductance $G$ as a function of $V_{gate}$ under 4 K with $V_{sd}$ = 5 mV. The inset shows an AFM friction image of the zGNR FET channel. The length of the zGNR in channel is ~50 nm. The scale bar in the inset is 50 nm. (c) Differential conductance *vs* $V_{sd}$ and $V_{gate}$ recorded at 4 K. The black arrow points to one of resonance peaks. (d) Differential conductance corresponding to $V_{gate}$ from 25 to 35 V. The black arrow points to one of resonance peaks.

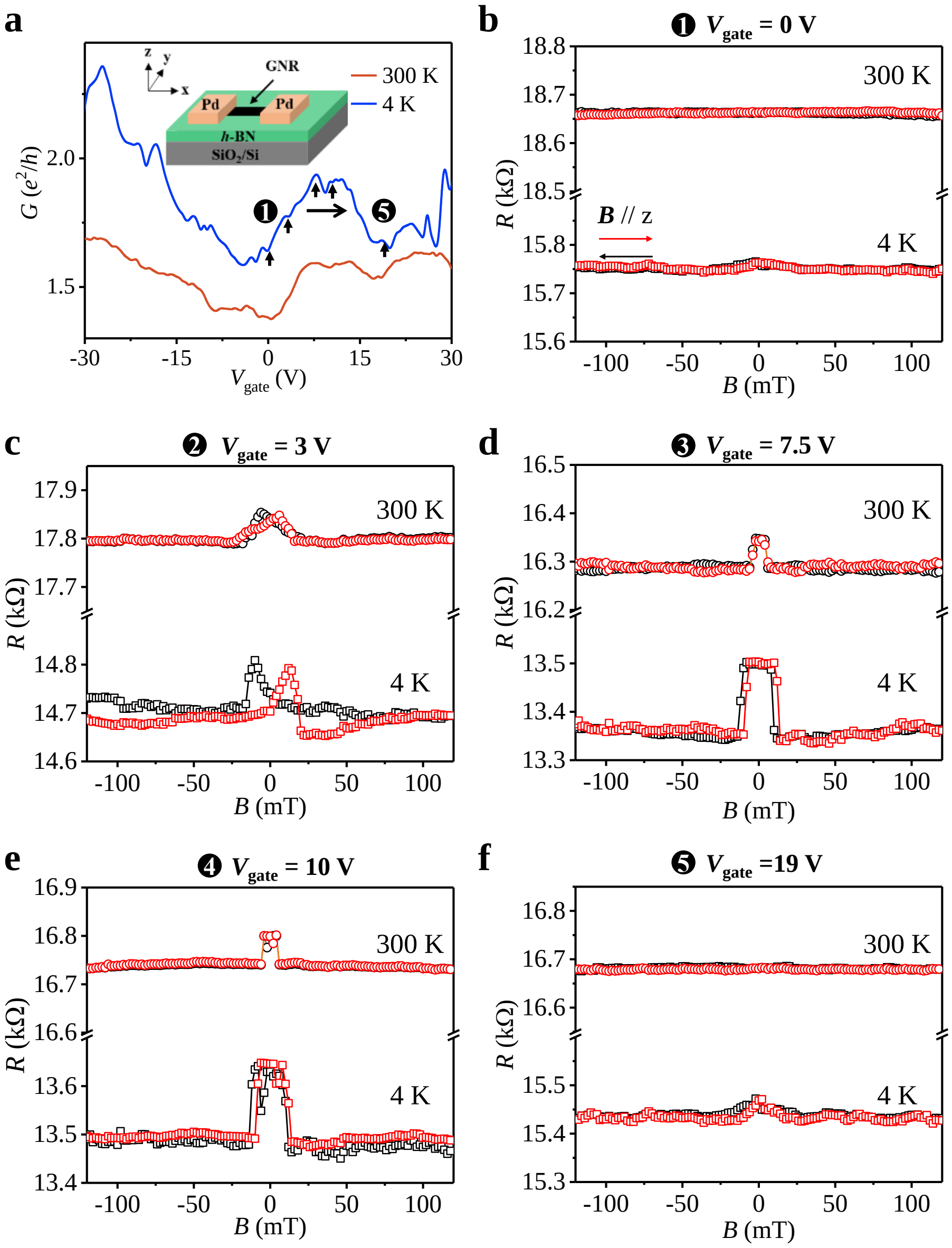

**Figure 3. Magneto-transport properties of the ~9 nm wide zGNR under different $V_{gate}$.** (a) Conductance $G$ as a function of $V_{gate}$ under 4 K and 300 K in the absence of magnetic field. (b-f) Magneto-resistance measured at 4 K and 300 K under different $V_{gate}$ with $V_{sd}$ = 5 mV. Here, the magnetic field $B$ is parallel to the direction of "z axis". The red (black) arrow represents the forward (backward) sweeping direction.

To probe the magnetic order of the edge states in zGNR, magneto-transport measurement was subsequently carried out in the zGNR transistor to characterize the magnetoresistance (MR) as a function of back-gate voltage ($V_{gate}$), magnetic field, source-drain bias ($V_{sd}$) and temperature. Figure 3a plots the $G$ *vs* $V_{gate}$ curves obtained at 4 K and 300 K, respectively. As can be seen, 5 test points of $V_g$ (0, 3, 7.5, 10 and 19 V) near the pronounced conductance peak are selected as shown in black arrows. Accordingly, we swept an out-of-plane magnetic field ($B$//z) with $V_{ds}$ = 5 mV at these $V_g$, and the results are presented in figures 3b-f. It is obvious that the shape of magnetoresistance curves depends on the applied $V_{gate}$. In particular, as shown in figures 3b & 3f, when $V_{gate}$ is tuned away from the regime of pronounced conductance peak, no hysteresis is observed in the response, which may correspond to the weak localization that is typically observed in graphene.[39] In sharp contrast, a clear hysteresis appears in the range of around ± 20 mT when $V_{gate}$ is tuned to near the broadened conductance peak (from 3 V to 15 V). Moreover, the resistance of the device always remains high at small magnetic fields, and switches suddenly to a low value once the fields exceed above 20 mT. Both observations coincide with the typical features of magnetic order reversal, and motivate us to take a closer look of these magneto responses.

During the field sweeping at $V_{gate}$ = 3 V, the resistance is gradually increased and followed by a steep drop to a stable value (see figure 3c). This response resembles the typical magnetization reversal behavior of a ferromagnet, which could be qualitatively attributed to the effect of anisotropic magnetoresistance (AMR). As predicted previously, intrinsic net moments could originate from *s*–*p* electrons at the two edges of zGNR. In a macro-spin view, it can be understood that within one edge, these net moments are paralleled aligned to form a uniform magnetization with magneto-crystalline anisotropy along the out-of-plane direction. While the magnetizations between the two edges are antiferromagnetically coupled that can only be rotated at a magnetic field exceeding the coupling strength. To further validate the explanation, we performed additional field sweeping transport measurements at 4 K while changing the direction to either *x*- or *y*-axis (see **Fig. S3**). The disappearance of discernable hysteresis in both responses supports the existence of magnetism as well as an out-of-plane anisotropy in the zGNR's edge state. In this case, the two edge state magnetizations are rotated in the same way by the moderate external field, so that their antiparallel configuration is always maintained during the whole sweep. Therefore, zGNR could be treated equivalently as one ferromagnet, which naturally results in the AMR-like curve. Similar MR responses are also found when $V_{gate}$ is set to 15 V (See **Fig. S9f** and **S10f**).

On the other hand, at $V_{gate}$ = 7.5 and 10 V (figures 3d and 3f), the MR curve exhibits a plateau-like shape where the resistance jumps between two stable values. The change of resistance ($\Delta R$) is ~60 Ω at 300 K and ~175 Ω at 4 K, corresponding to a MR ratio of ~0.5 % and ~1.3 %, respectively. There are a few reasons to account for the formation of plateau-like feature. Different from the case of $V_{gate}$ = 3 V, the origin of the $\Delta R$ at $V_{gate}$ = 7.5 and 10 V seems more complicated. With the possible lowering of coupling strength between the edge states, the magnetic field could now cause a spin flop of antiferromagnetically coupled magnetizations. In this case, the antiparallel magnetization configuration becomes parallel with the increasing of fields. Similar to the giant magnetoresistance (GMR) effect observed in antiferromagnetic spin valve systems, the transition between the two configurations leads to the plateau-like shape in the MR responses.[40] Such a continuous tuning of Fermi level via $V_{gate}$ opens up opportunities for probing magnetic ordering in zGNR, and sheds light on potential voltage-controlled magnetic devices.

It is also interesting to mention that the signature of sizable MR extends to even 350 K, which is the highest temperature available in the cryostat. We believe that the magnetoresistance could be related to the type of *s*–*p* electron magnetism in the zGNR.[8,41,42] When magneto-transport measurement was carried out with magnetic field parallel to current (**Fig. S3a**) and in the plane of the substrate but perpendicular to the current (**Fig. S3b**) at 4K under different $V_{gate}$, no obvious switching-like hysteresis was observed in them. It clearly shows the magnetic anisotropy of zGNR. It indicates that the anisotropy of zGNR has not in-plane component. As steep switching of magnetoresistance was only observed with the out-of-plane magnetic field, zGNRs embedded in *h*-BN are of an out-of-plane direction magneto-crystalline anisotropy. In addition, the flipping of magnetic moments within one edge state could also leads to the steep switch as the spin polarized edge states have no in-plane components. The results are in agreement with previous predictions.[2]

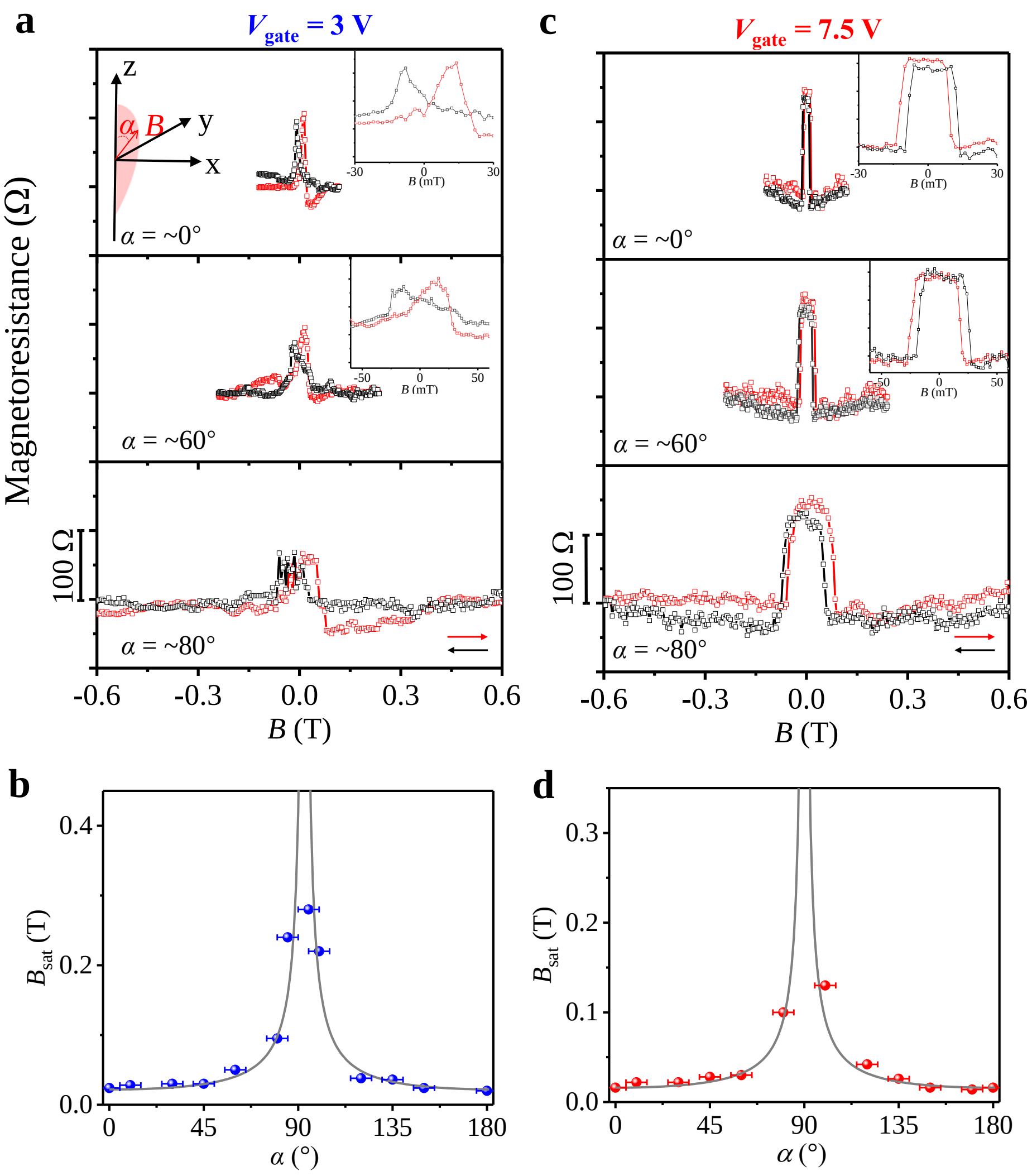

**Figure 4. Angle dependence of magnetoresistance under different $V_{gate}$ at 4K.** (a) Magnetoresistance curves at different *α*, where *α* is the separation angle between z-axis and the magnetic field *B* in the z-y plane, when $V_{gate}$ = 3 V. Insets in the top and middle panels show the hysteresis of magnetoresistance in the zoom-in magnetic field

regime. (b) Extracted saturation field ($B_{sat}$) as a function of $\alpha$. Note that there is a small deviation (±5°) in determining the separation angle because of our manual rotation on the sample holder in the experimental set-up. (c) Magnetoresistance hysteresis at different $\alpha$ when $V_{gate}$ = 7.5 V. (d) Extracted saturation field ($B_{sat}$) as a function of $\alpha$.

Next, we proceed to investigate the switching process of the edge states' magnetization by measuring the MR with different tilting angle ($\alpha$) between the applied magnetic field $B$ and $z$-axis in the $y$-$z$ plane. Figure 4a and 4c shows the MR hysteresis measured with $\alpha$ = ~0°, ~60° and ~80° when $V_{gate}$ = 3 V and $V_{gate}$ = 7.5 V at 4 K, respectively. Results of more tilting angles in the $y$-$z$ plane are shown in **Fig. S4 and S7.** It is evident that it requires a significantly larger magnetic field to complete magnetization switching for a large $\alpha$. To discuss this more quantitatively, the effective saturation field $B_{sat}$ is defined as the field where the magnetization switching completes, that is, the field at which the hysteresis disappears. Figure 4b shows that $B_{sat}$ rises sharply as the tilt angle approaches 90°. The phenomenon is quite similar to the characteristic of magnetization switching associated with domain walls[43] in many material systems[44-46]. It is found that the depinning field $B_{dep}(\alpha)$ fits the formula $B_{dep}(\alpha) = B_0/\cos\alpha$ very well, where $B_0$ is the depinning field along the easy axis. The good fit to the experimental results indicates that the dominant switching mechanism could be domain-wall motion. Similarly, Figure 4c shows the MR hysteresis measured with three tilt angles between $B$ and y-axis in the y-z plane when $V_{gate}$ = 7.5 V at 4 K. Details about all angles measured are shown in **Fig. S7**. Figure 3d shows that the experimental data of $B_{sat}$ are also in a good fit to the formula $B_{dep}(\alpha) = B_0/\cos\alpha$. Similar results are observed in the MR hysteresis measured with tilting angles between $B$ and the zGNR (also x-axis) in the x-z plane when $V_{gate}$ = 3 V and 7.5 V at 4 K. It is noted that MR change ($\Delta R$) and hysteresis are not observed in the MR measured with different tilt angle between $B$ and x-axis in the x-y plane. All these additional results further verify that the magnetic easy axis of zGNR is along the out-of-plane direction, and the switching of edge state magnetization is achieved via the domain pinning and domain wall motion. The results of MR hysteresis measured with different tilt angle between $B$ and x-axis in the x-z plane when $V_{gate}$ = 3 V (**Fig. S5**) and $V_{gate}$ = 7.5 V (**Fig. S8**) at 4 K also support out-of-plane magnetic anisotropy of zGNRs embedded in $h$-BN.

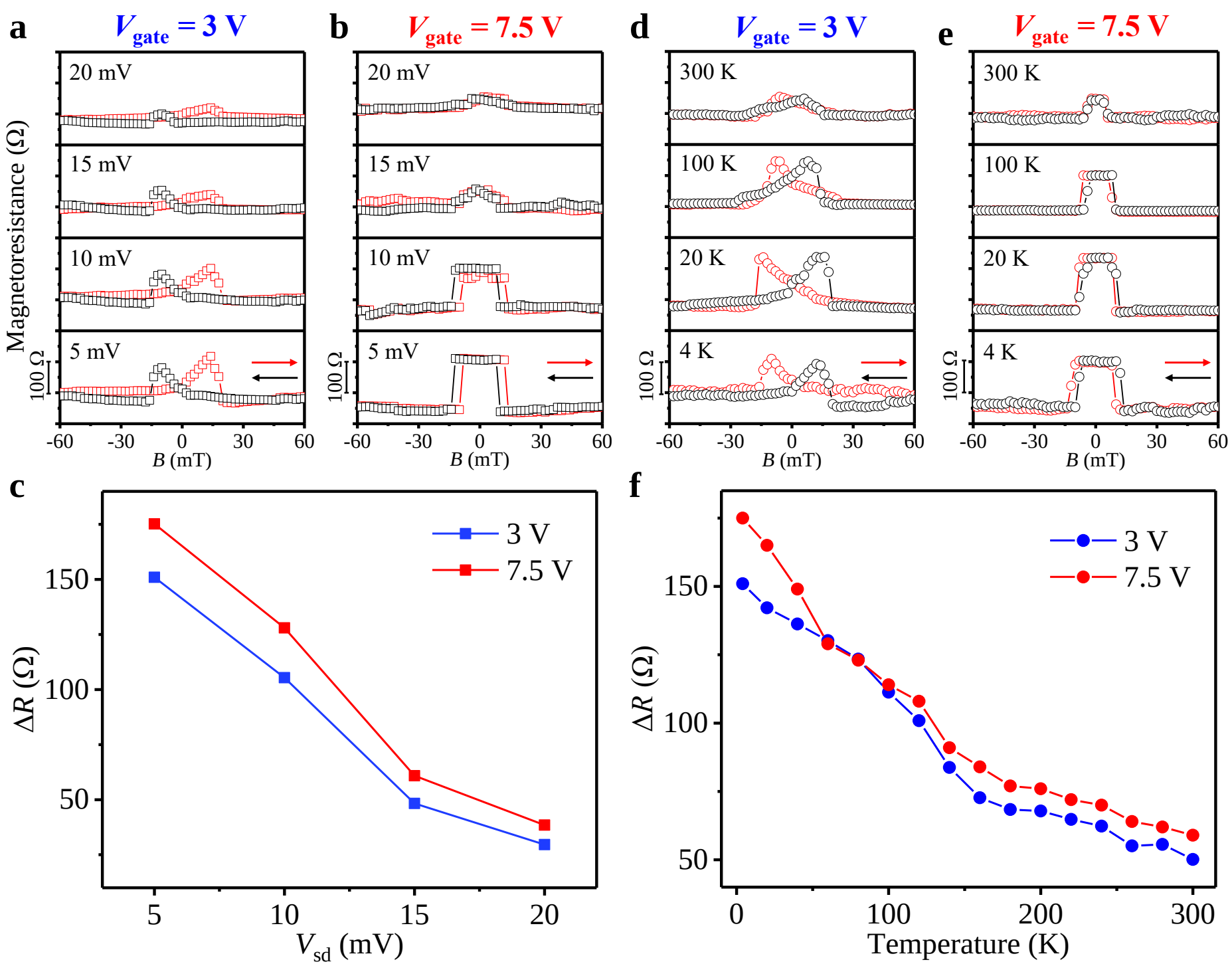

**Figure 5. Bias dependent magnetoresistance of the ~9nm wide zGNR field effect transistor at $T$ = 4K.** Magnetoresistance change at $V_{sd}$ = 5 mV, 10 mV, 15 mV and 20 mV under (a) $V_{gate}$ = 3V and (b) $V_{gate}$ = 7.5V. The red (black) arrow represents the sweep direction of the magnetic field from -0.06T (0.06T) to –0.06 T (0.06 T). (c) $V_{sd}$ dependence of magnetoresistance change ($\Delta R$). $\Delta R$ is calculated by subtracting a polynomial fitting from the device resistance.

The dependence of the magnetoresistance changes ($\Delta R$) on the bias voltage and the temperature are also investigated. The intensity of the resistance jump in this device exhibits a dependence on $V_{sd}$, as shown in Figure 5a-c (also see **Fig. S9** and **S10**). It is found $\Delta R$ decreases with the increase of $V_{sd}$ in both types of magnetoresistance changes, and the signal becomes invisible above 20 mV. It is also noted that the width of magnetic field for $\Delta R$ decreases with $V_{sd}$. They may be related to a slightly non-linear charge injection at low biases[47] and the realignment of the DOS along zGNR under large bias which overcomes the exchange splitting in zGNR, its physical origin needs further investigation. In addition, the $\Delta R$ shows a dependence on the temperature (See in figure 5d-f, **Fig. S11** and **S12**). It decreases with the raising of temperature. The thermal vibration of lattices could cause more spin-dependent scattering and then suppress the magnetic response of the zGNR device.

Similar magnetoresistance responses are observed in three different samples with comparable coercive field values. Such hysteretic behaviors are the hallmarks of ferromagnetism. We believe that the hysteresis and switch of magnetoresistance could be caused from the magnetic edge states in zGNR.

**Discussion**
The observed magnetism in the embedded zGNRs corroborate the nature of the predicted emergent magnetic order in zGNRs. The results are not only of scientific interest in the interplay between electric fields and electronic spin degree of freedom in solids[6,7,48,49] but may also open a new path to explore spintronics[3,50-52]at the nanometre scale, based on graphene. Further, first-principle calculations showed that zigzag-oriented GNRs could display the property of half-metallicity when electric fields were applied across the zigzag directions[4,10]. As a spin-polarized conductance mode, the half-metallicity is an important enabling phenomenon for the development of graphene-based spintronic devices because it provides an energy-efficient route to control magnetic properties using electric field. It is expect that the availability of zGNRs will enable the characterization of their predicted spin-related properties, such as spin confinement[11,53] and filtering[54], and will ultimately add the spin degree of freedom to graphene-based circuitry. Therefore, our work ignites the potential of structure engineering for driving nonmagnetic graphene into magnetic phases, thereby opening new directions for exploring spin transport in graphene-based magnets.

## Methods

**Etching process on h-BN and GNR growth**
Firstly, *h*-BN flakes from its bulk were mechanically cleaved onto quartz substrates, followed by a process of annealing at 650 °C in an $O_2$ flow for 60 min to remove organic residues on the surface. Next, a $ZnCl_2$ solution was spun at 4,000 r.p.m. for 100 s onto the substrate and then baked for 10 min. Etching was carried out at a furnace under $H_2$ atmosphere at high temperature.[20] After etching, nano-trenches form on *h*-BN, the quartz substrates with *h*-BN flakes were subjected to ultrasonic bathing for cleaning in an HCl solution, deionized water and acetone, in sequence. Subsequently, the quartz substrates were loaded into another furnace and heated to 1,280 °C in an argon flow, and after that, a $C_2H_2$ flow and a mixture of silane/argon (5% mole ratio of silane) were introduced into the chamber for zGNR growth.[22,23]

**SNVM measurement**
Both typical SNVM working modes were performed. The ISO-B mode needs a fixed the microwave (MW) frequency at which the NV is resonant in the bias field.[55] The scanning results show the contour of a magnetic field along the NV axis. While the other: Full-B mode, records the full-range ODMR spectrum of NV centers during MW frequency sweeping, giving the ability to acquire the quantitative stray field magnitude along the NV center. In figure 1c, by applying an external vertical magnetic field, the ODMR splits into two peaks, whose corresponding MW frequency is 2805 MHz and 2934 MHz respectively. To obtain a higher sensitivity, the MW frequency was fixed at 2801 MHz for ISO-B mode scanning.

**Device fabrication**
Short channel transistors based on zGNR in two-terminal configuration were fabricated here. Electrodes with ~50 nm thick Pd layer was firstly applied onto the zGNR. After that, another 10 nm thick Pd layer was evaporated after rotating the substrate for ~30 degree, and then an ultra-short channel between electrodes was fabricated by self-shadowing. The heavily *p*-doped Si substrate serves as the global back gate with the gate dielectric of 285 nm $SiO_2$. The thickness of the *h*-BN flake is about 20 nm.

**Transport measurements**

Transport measurement at low temperature was carried out in a Quantum Design cryo-magnetic PPMS system with a base temperature of ~2 K. Two Keithley2400 DC sources serve as voltage supply for $V_{sd}$ and $V_{gate}$ and measured the differential conductance using a lock-in amplifier (Stanford Research Systems SR830). Temperature and $V_{ds}$ dependence of magnetoresistance were measured using a Keithley Sourcemeter 4200. The source-drain voltage ($V_{ds}$) was varied between 5 mV and 20 mV. Transport measurements were carried out in a two-terminal configuration.

**SQUID magnetometry measurements**

Since SQUID magnetometry is an integral technique, special care was taken with sample preparation for SQUID measurements. The sample holder used in SQUID measurements is made of a weakly paramagnetic material. Sample holders were cleaned and subjected to SQUID measurement to ensure that no ferromagnetic contamination is involved. Raw SQUID signals shown in Figure S2 includes the paramagnetic contribution of the sample holder. Several *h*-BN flakes were pealed from the growth substrate and transferred on the 300 nm $SiO_2$/Si substrate to order to avoid the influence of carbon grown on the growth substrate (normally Quartz). M(H) curves were measured at 350 K and 10 K with the in-plane magnetic field ranging from -30 mT to 30 mT. Magnetic field offset is less than 2E-5 T (i.e., 0.007% of the total measurement range, which is negligibly small in both the measurement of the out-of-plane magnetization loops and the case of the in-plane measurements, and therefore no compensation was carried out.

**Contributions**

H.W. and H.S.W. directed and supervised the research work. H.W. conceived and designed the research. C.J., H.S.W. and C.C. contributed equally to this work. H.S.W., C.C. and X.W. fabricated the zGNR devices and carried out transport measurements. C.J., L.C., C.C., Y.W., Y.F. and Z.X. performed the growth of zGNRs. H.S.W., C.C., C.J. and Z.K. performed AFM and SEM measurements. H.S.W., Y.L. and G.M. performed MPMS measurements. H.W., H.S.W., and Y.Y. analyzed the experimental data and wrote the manuscript with the contribution from all authors. All the authors contributed to critical discussions of the results and manuscript.

**Acknowledgments**

H.W. thanks C.G. Duan (East China Normal University), D. Sun and W. Han (Peking University) for helpful discussion. H.W. thanks M.S. Guo and A.M. Tong from CIQTEK for the NV center measurements (Diamond III QDAFM). This work was

supported by the National Natural Science Foundation of China (91964102, 51772317, 12004406, 62074099, 12304113), the National Key R&D Program of China (2023YFF0612502, 2022YFF0609800, 2017YFF0206106), Shanghai Collaborative Innovation Project (XTCX-KJ-2024-02), the Strategic Priority Research Program of Chinese Academy of Sciences (XDB30000000), the Science and Technology Commission of Shanghai Municipality (20DZ2203600), Shanghai Post-doctoral Excellence Program (2021515), China Postdoctoral Science Foundation (BX2021331, 2021M703338, 2021M693425, 2021K224B), ShanghaiTech Soft Matter Nanofab (SMN180827) and ShanghaiTech Material and Device Lab. K.W. and T.T. acknowledge support from JSPS KAKENHI (19H05790, 20H00354, 21H05233) and A3 Foresight by JSPS.

# Supplementary Information

## Signatures of magnetism in zigzag graphene nanoribbon embedded in *h*-BN lattice

Chengxin Jiang,[1,#] Hui Shan Wang,[1,2,3,#*] Chen Chen,[1,#] Lingxiu Chen,[4] Xiujun Wang,[1,2] Yibo Wang,[1] Ziqiang Kong,[1,2] Yuhan Feng,[1,2] Yixin Liu,[1] Yu Feng,[1,2] Chenxi Liu,[1,2] Yu Zhang,[1,2] Zhipeng Wei,[5] Maosen Guo,[6] Aomei Tong,[6] Gang Mu,[1,2] Yumeng Yang,[7] Kenji Watanabe,[8] Takashi Taniguchi,[9] Wangzhou Shi,[3] Haomin Wang[1,2*]

[1] State Key Laboratory of Functional Materials for Informatics, Shanghai Institute of Microsystem and Information Technology, Chinese Academy of Sciences, Shanghai 200050, China.
[2] Center of Materials Science and Optoelectronics Engineering, University of Chinese Academy of Sciences, Beijing 100049, China.
[3] Key Laboratory of Optoelectronic Material and Device, Department of Physics, Shanghai Normal University, Shanghai 200234, China.
[4] School of Materials Science and Physics, China University of Mining and Technology, Xuzhou 221116, China.
[5] State Key Laboratory of High Power Semiconductor Lasers, Changchun University of Science and Technology, Changchun, 130022, People's Republic of China.
[6] CIQTEK Co,. Ltd., Hefei 230000, China.
[7] School of Information Science and Technology, ShanghaiTech University, Shanghai 201210, China.
[8] Research Center for Functional Materials, National Institute for Materials Science, 1-1 Namiki, Tsukuba, 305-0044 Japan.
[9] International Center for Materials Nanoarchitectonics, National Institute for Materials Science, 1-1 Namiki, Tsukuba 305-0044, Japan.

E-mail: hswang2024@shnu.edu.cn; hmwang@mail.sim.ac.cn

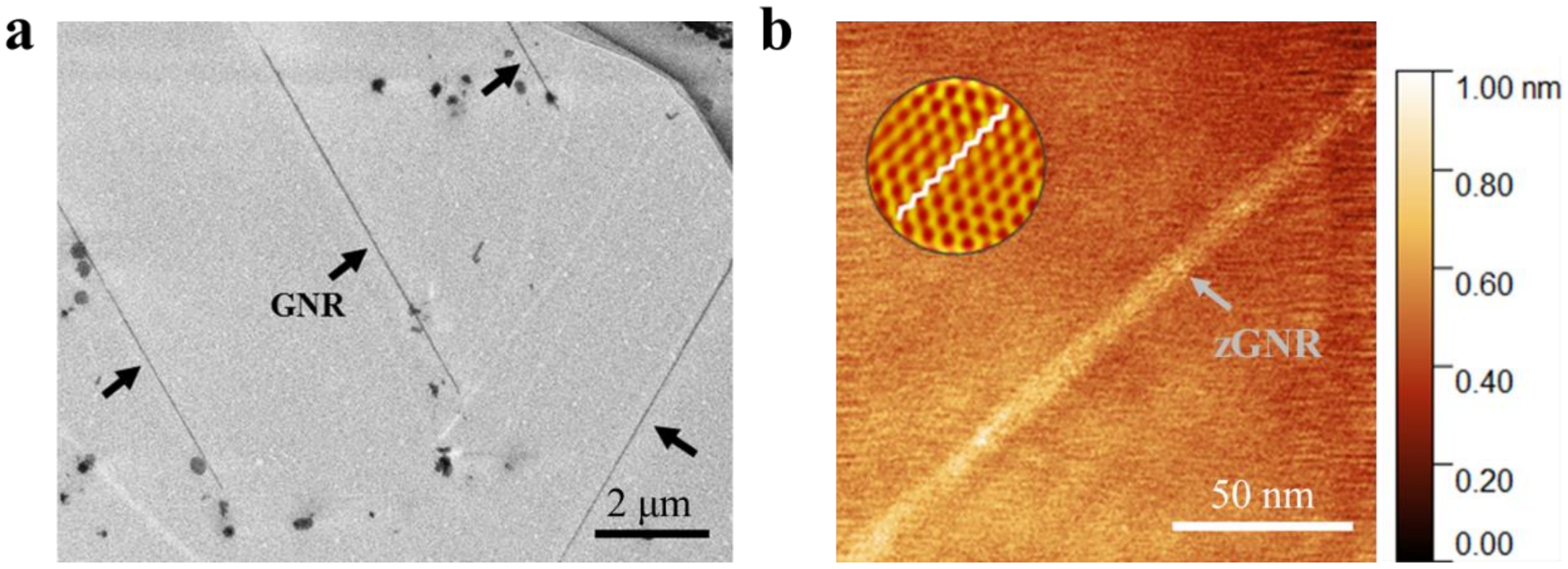

**Figure S1. Characterization of zGNR embedded in *h*-BN. (a)** A SEM image of zGNRs embedded in *h*-BN. The black arrows point to zGNRs. Scale bar is 2 µm. **(b)** AFM height image of the zGNR embedded in *h*-BN. The grey arrow points to the zGNR. Scale bar is 50 nm. The inset shows the lattice resolution friction image of *h*-BN which indicates the orientation of GNR.

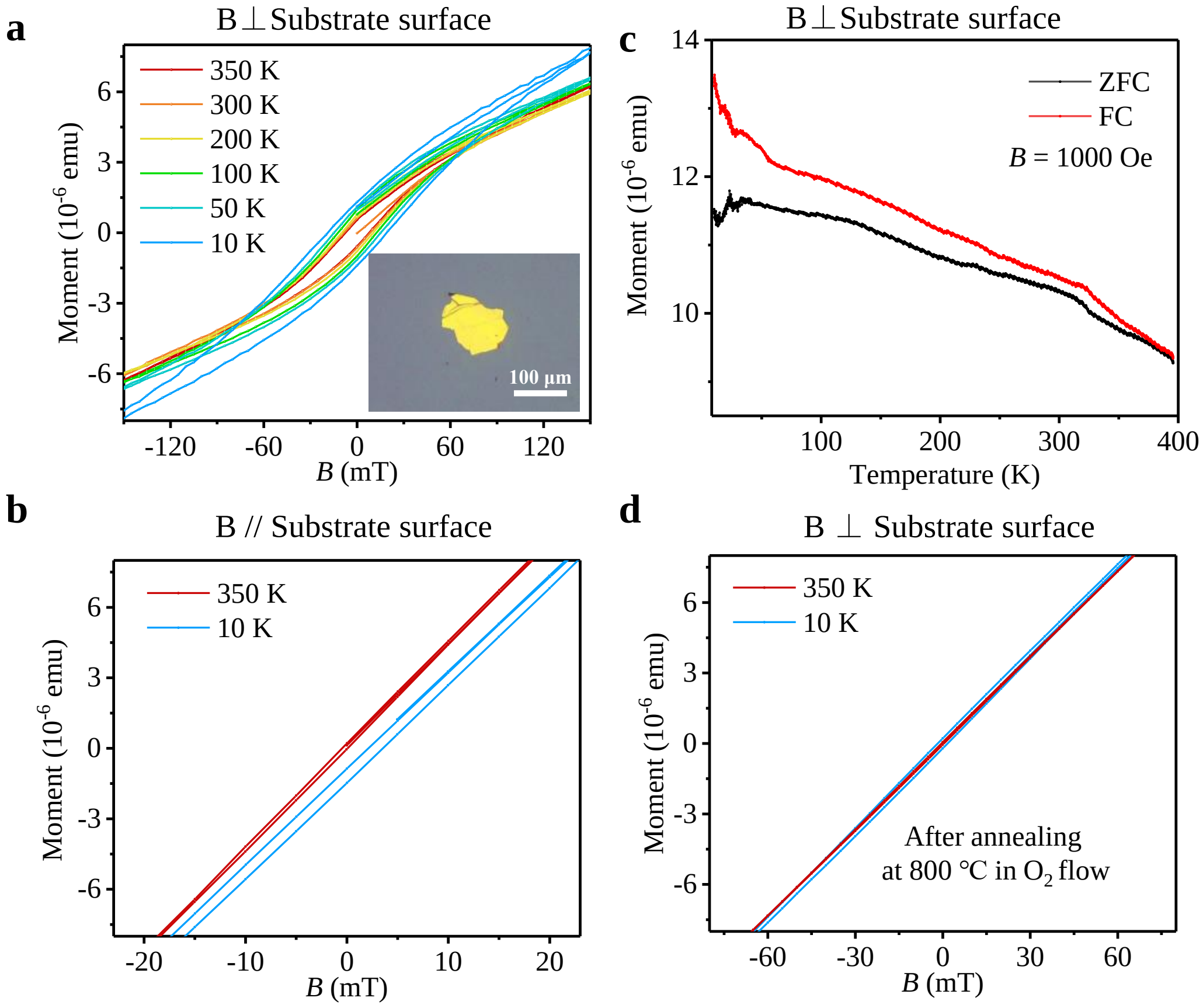

**Figure S2. Macroscopic investigation in magnetic ordering in zGNR.** Raw SQUID data of zGNRs (a) with the direction of magnetic field *B* perpendicular to the surface of substrate and (b) with *B* // substrate surface. Inset shows an optical image of the *h*BN flake with zGNRs on a 300 nm $SiO_2$/silicon substrate. (c) ZFC and FC *M*-*T* curves measured from 10 to 395 K with *B*⊥the surface of substrate under the applied field = 0.1 T. (d) Raw SQUID data of the same sample with *B*⊥the surface of substrate after the sample was subject to annealing at 800 °C in a $O_2$ flow to burn out all zGNRs. The variation in the slopes of M-H loop originate from a small change of sample position in cryostat as it is very difficult to place the sample at the exactly same position.

## Macroscopic magnetic response of zGNRs

The magnetic responses of *h*-BN embedded with zGNRs were measured by superconducting quantum interference device-vibrating sample magnetometer (SQUID-VSM) MPMS2 from Quantum Design. Figure S2a provides raw data of out-of-plane SQUID measurements at different temperatures while Figure S2b shows in-plane responses of the same sample to applied external fields at different temperature. Obvious *M-H* hysteresis loop only was observed for out-of-plane magnetic responses. A coercivity ($B_C$) is 19.98±0.01 mT at 10 K and 9.97±0.01 mT at 350 K. It indicates that the sample with zGNRs exhibits a ferromagnetic behavior even at room temperature and magnetic anisotropy with the easy axis along the direction of out-of-plane. Shown in Fig. S2c are the dependences of mass magnetization on temperature K.

The two curves measured under zero-field cooling (ZFC) and field cooling (FC) conditions show different features. The observable divergence of the ZFC and FC data suggests that our sample is not purely paramagnetic at low temperature. Moreover, both FC and ZFC curves display a monotonic decrease of the magnetization with the increasing temperature. Combined with the fact that the apparent hysteresis loop behavior can be observed in *M–H* curves, the observable divergence of the ZFC and FC curves (Figure S2c) suggest the existence of ordered magnetism. In addition, the phenomenon of *M–H* hysteresis is absent in magnetic response of the sample where zGNRs were burn out (Figure S2d). It is clear that the origin of the weak ferromagnetic response is most likely connected to the zGNRs embedded in *h*-BN.

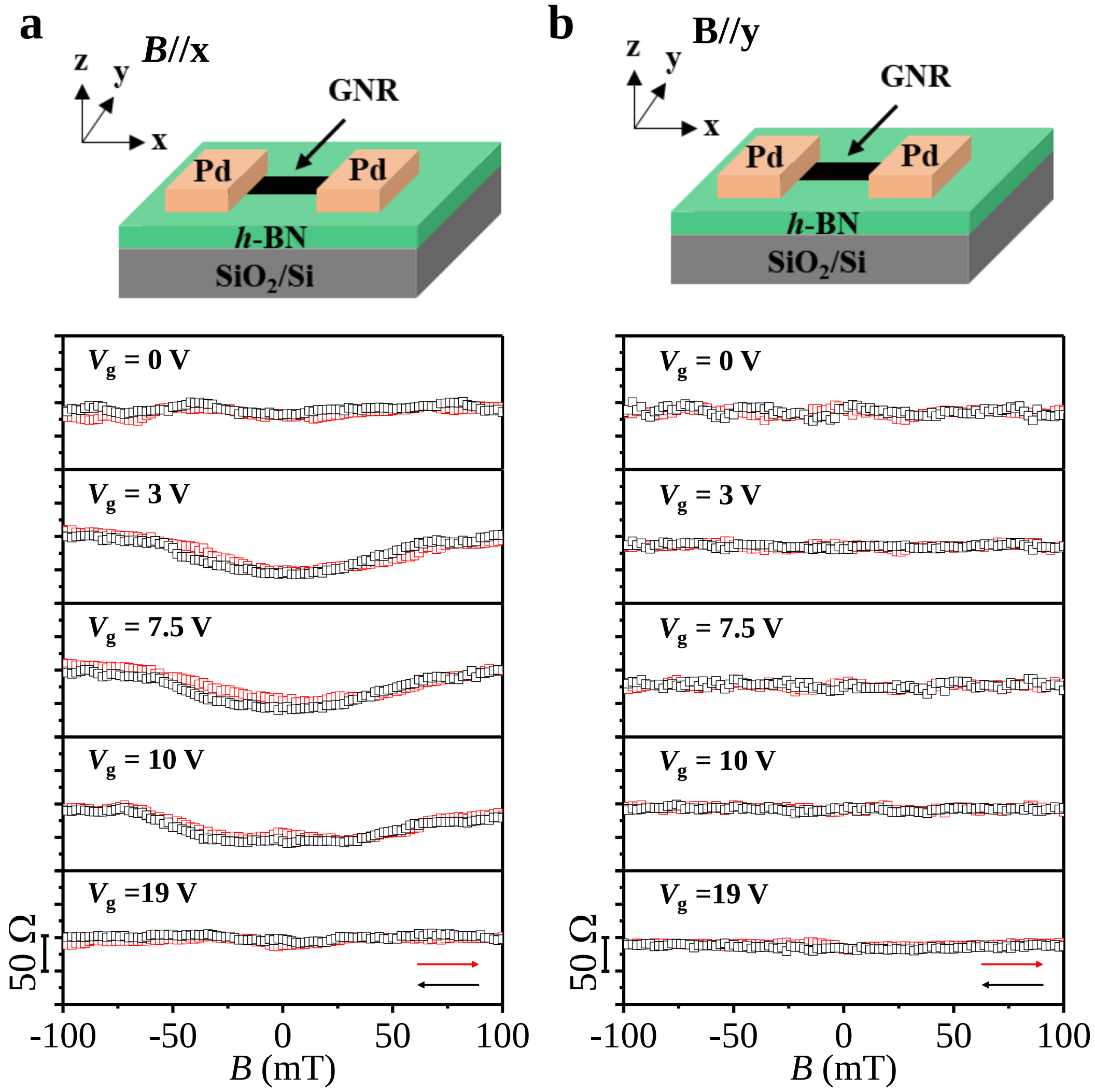

**Figure S3.** Magneto-resistance measured at 4K under different $V_{gate}$ with $V_{sd}$ = 5 mV when the magnetic field $B$ is parallel to the direction of (a) “x axis” and (b) “y axis”. The red (black) arrow represents the forward (backward) sweeping direction.

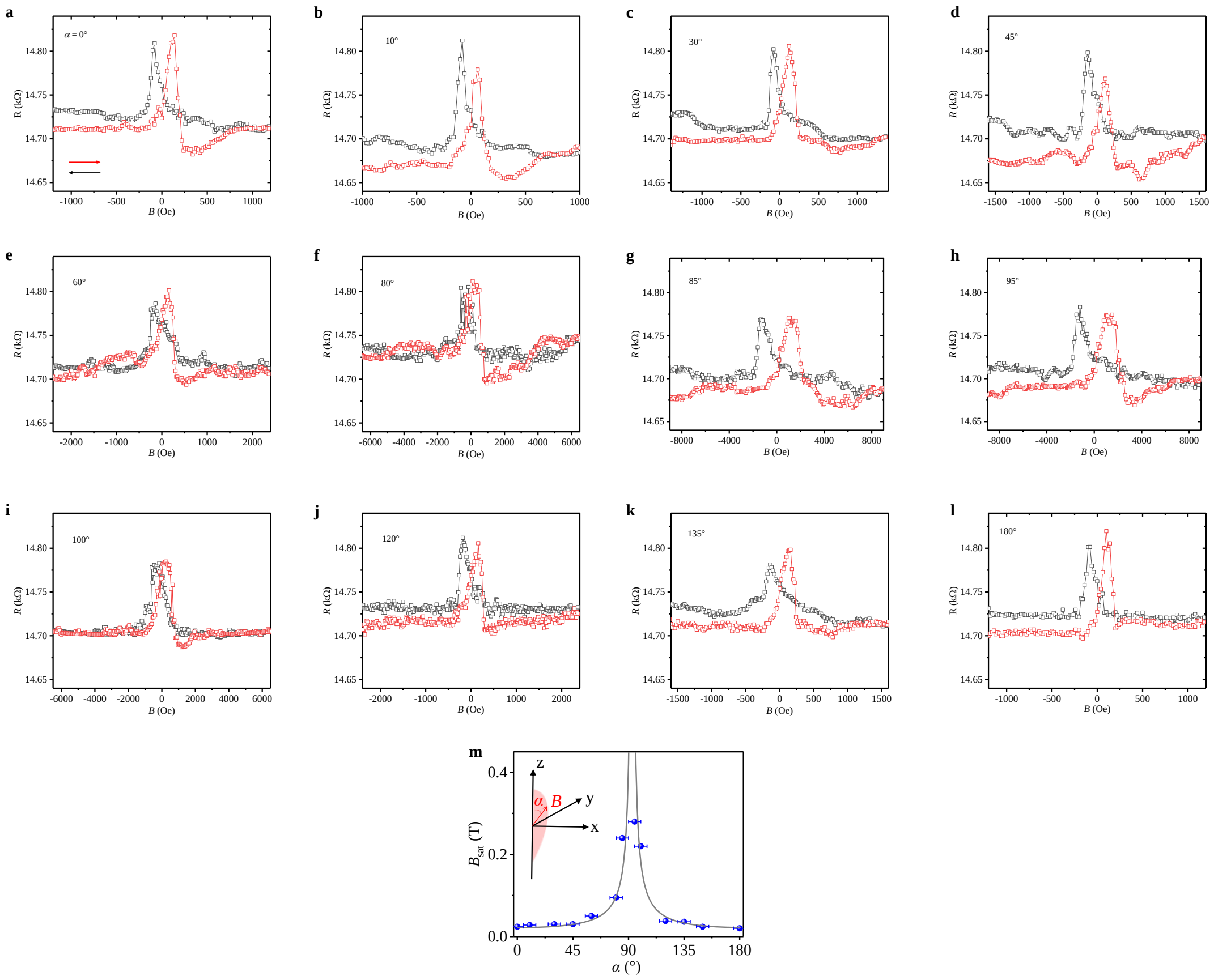

**Figure S4. Angle dependence of magnetoresistance under $V_{\text{gate}}$ = 3V at 4K.** (a-l) Magnetoresistance hysteresis curves at different $\alpha$, where $\alpha$ is the separation angle between z-axis and the magnetic field $B$ in the y-z plane. (m) Extracted saturation field ($B_{\text{sat}}$) as a function of $\alpha$. Note that there is a small deviation in determining the separation angle ( ±5°) because of our manual rotation on the sample holder in the experimental set-up.

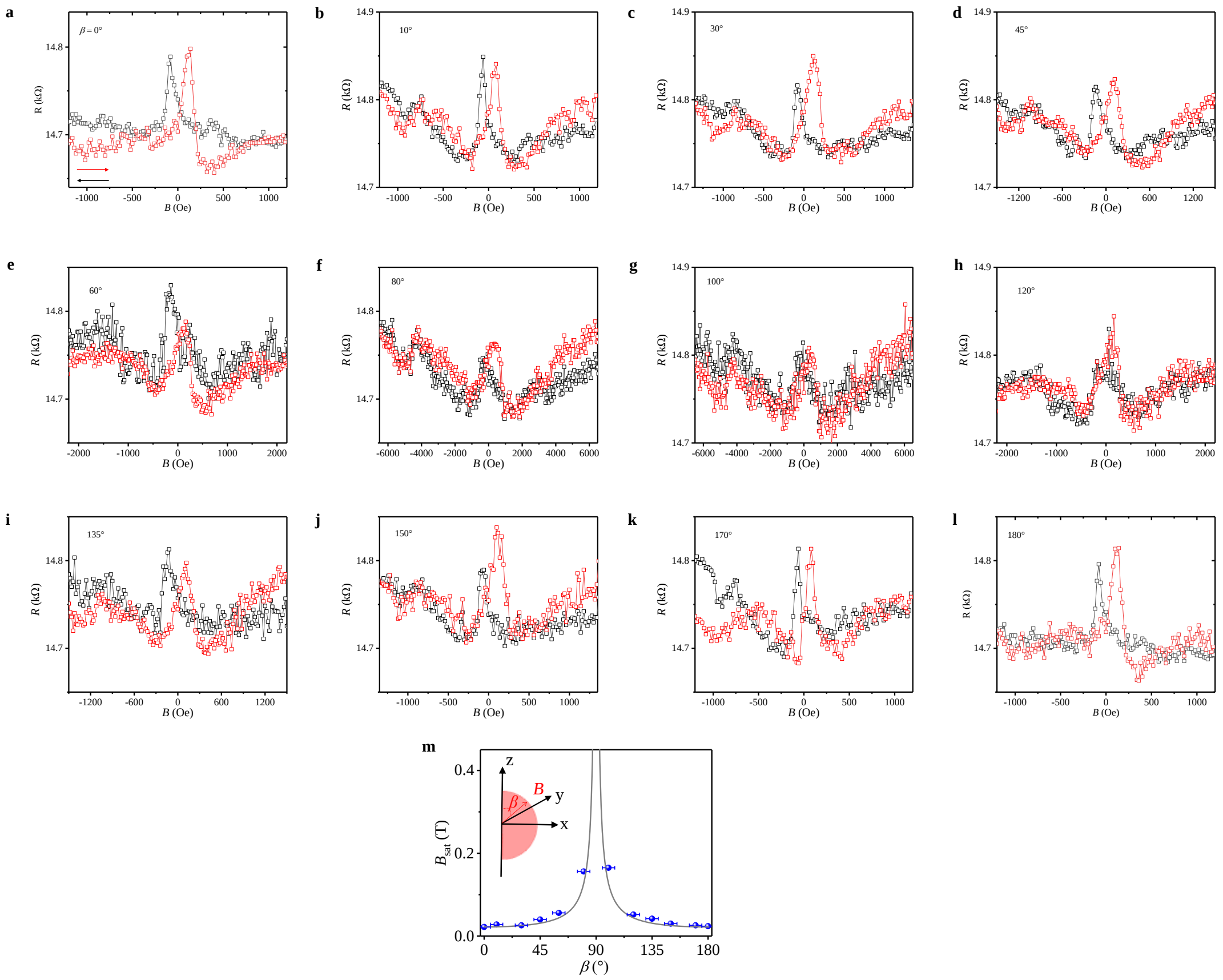

**Figure S5. Angle dependence of magnetoresistance in x-z plane under $V_{\mathrm{gate}}$ = 3 V at 4K.** (a-l) Magnetoresistance hysteresis curves at different $\beta$, where $\beta$ is the separation angle between z-axis and the magnetic field $B$ in the x-z plane. (m) Extracted saturation field ($B_{\mathrm{sat}}$) as a function of $\beta$. Note that there is a small deviation in determining the separation angle (±5°) because of our manual rotation on the sample holder in the experimental set-up.

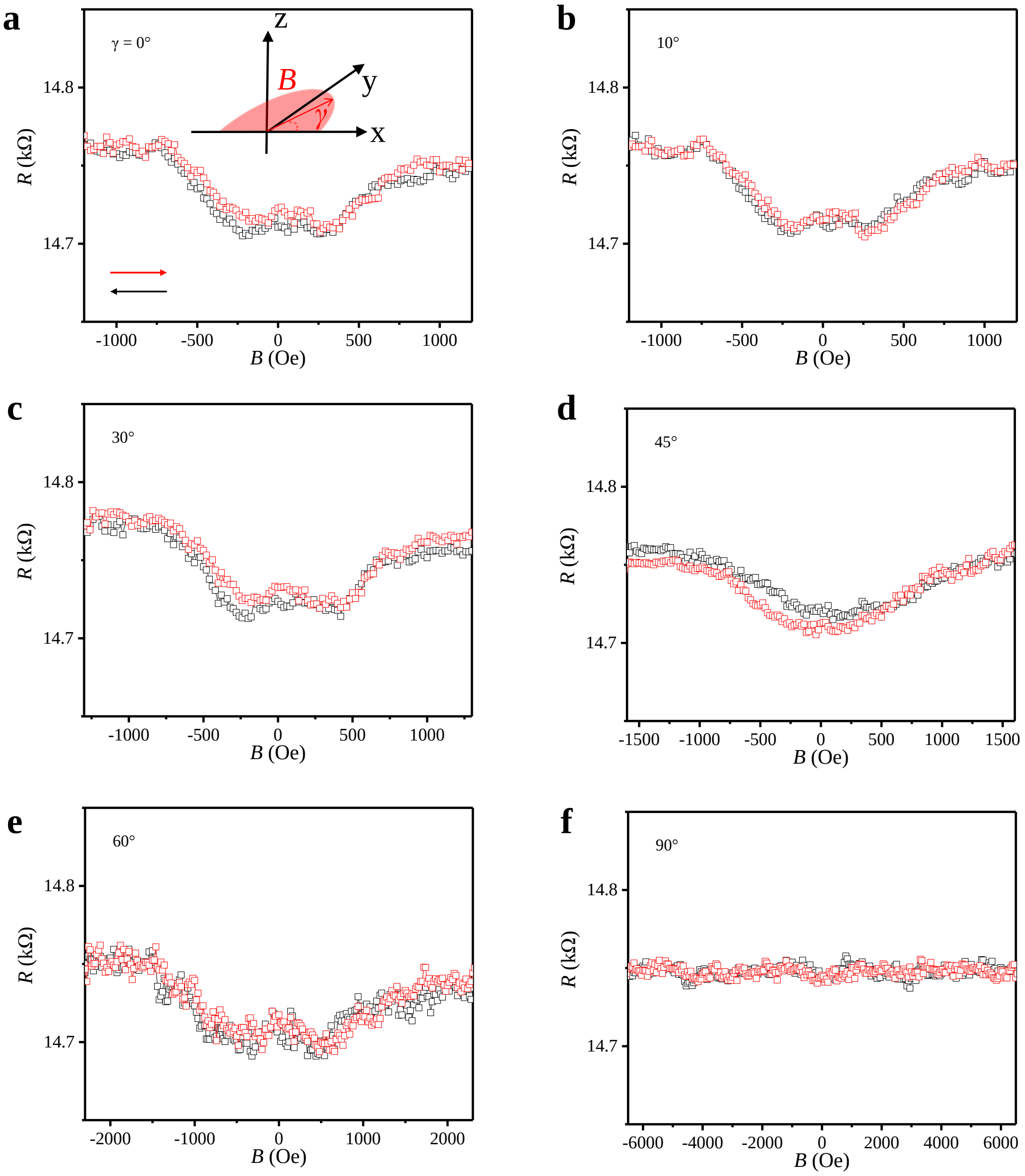

**Figure S6. Angle dependence of magnetoresistance in x-y plane under $V_{gate}$ = 3 V at 4K.** (a-f) Magnetoresistance hysteresis curves at different *γ*, where *γ* is the separation angle between x-axis and the magnetic field *B* in the x-y plane.

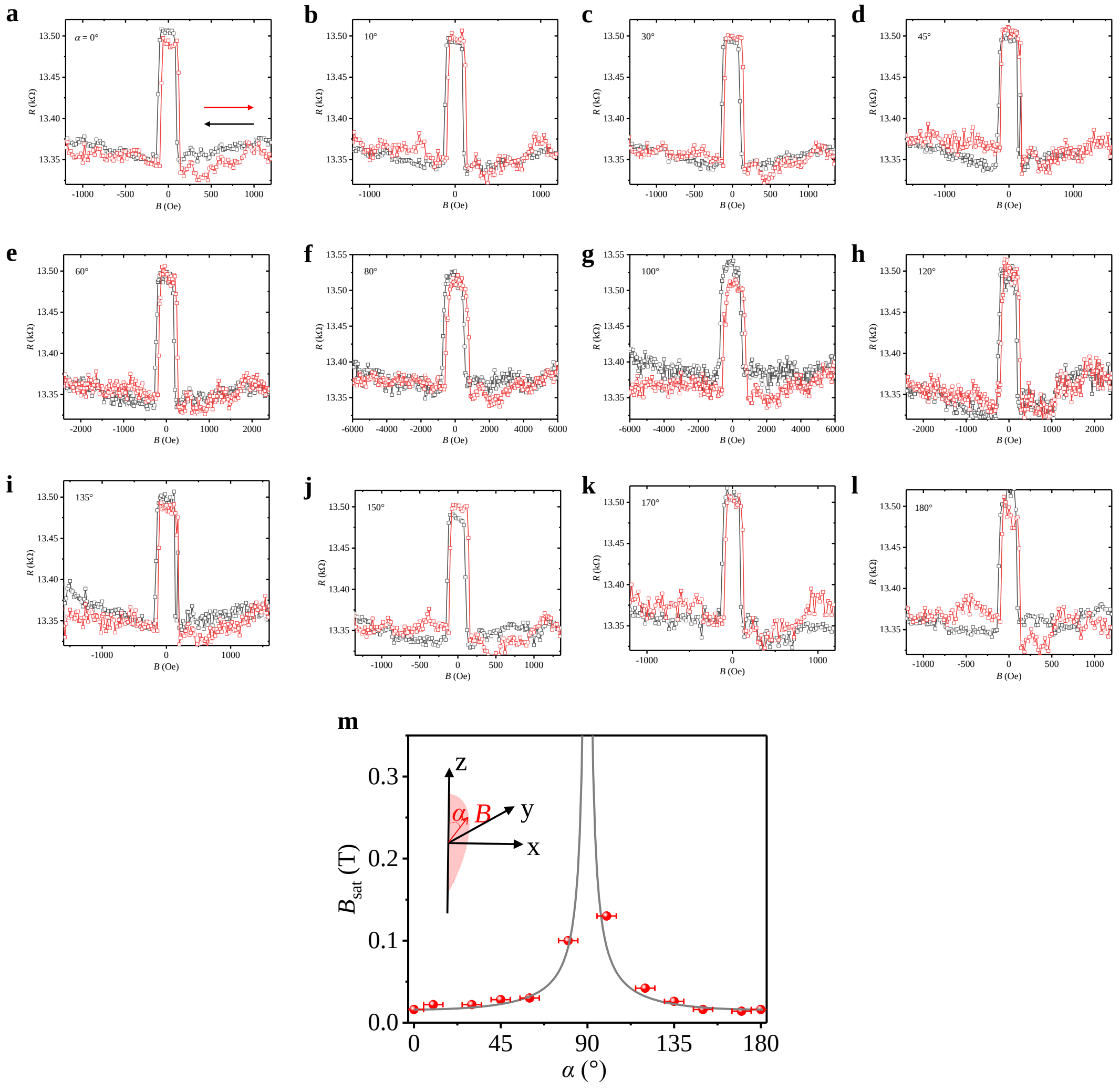

**Figure S7. Angle dependence of magnetoresistance in y-z plane under $V_{gate}$ = 7.5 V at 4 K.** (a-l) Magnetoresistance hysteresis curves at different $\alpha$, where $\alpha$ is the separation angle between z-axis and the magnetic field $B$ in the y-z plane. (m) Extracted saturation field ($B_{sat}$) as a function of $\alpha$. Note that there is a small deviation in determining the separation angle (±5°) because of our manual rotation on the sample holder in the experimental set-up.

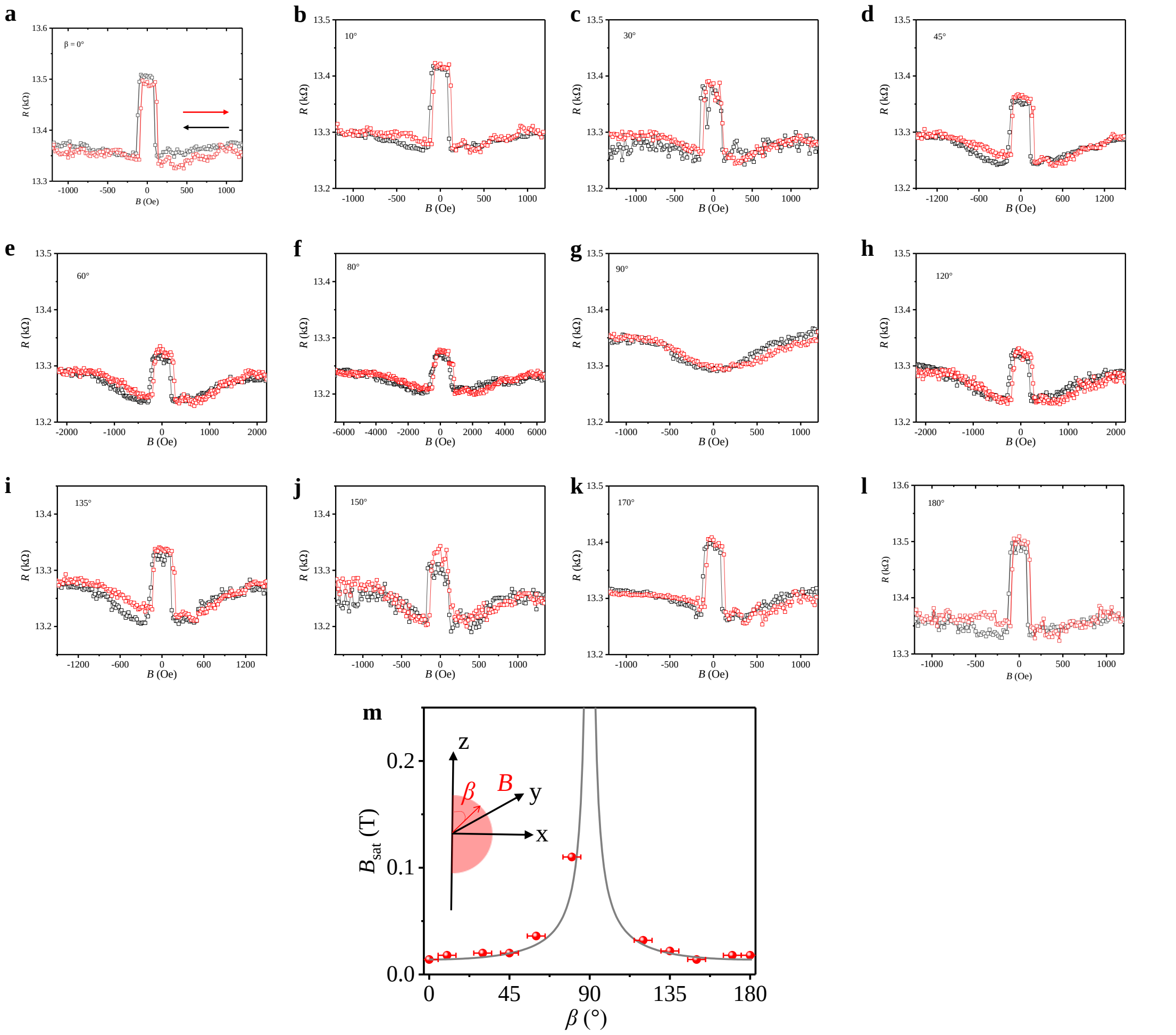

**Figure S8. Angle dependence of magnetoresistance under $V_{gate}$ = 7.5 V at 4 K.** (a-l) Magnetoresistance hysteresis curves at different $\beta$, where $\beta$ is the separation angle between z-axis and the magnetic field $B$ in the x-z plane. (m) Extracted saturation field ($B_{sat}$) as a function of $\beta$. Note that there is a small deviation in determining the separation angle (±5°) because of our manual rotation on the sample holder in the experimental set-up.

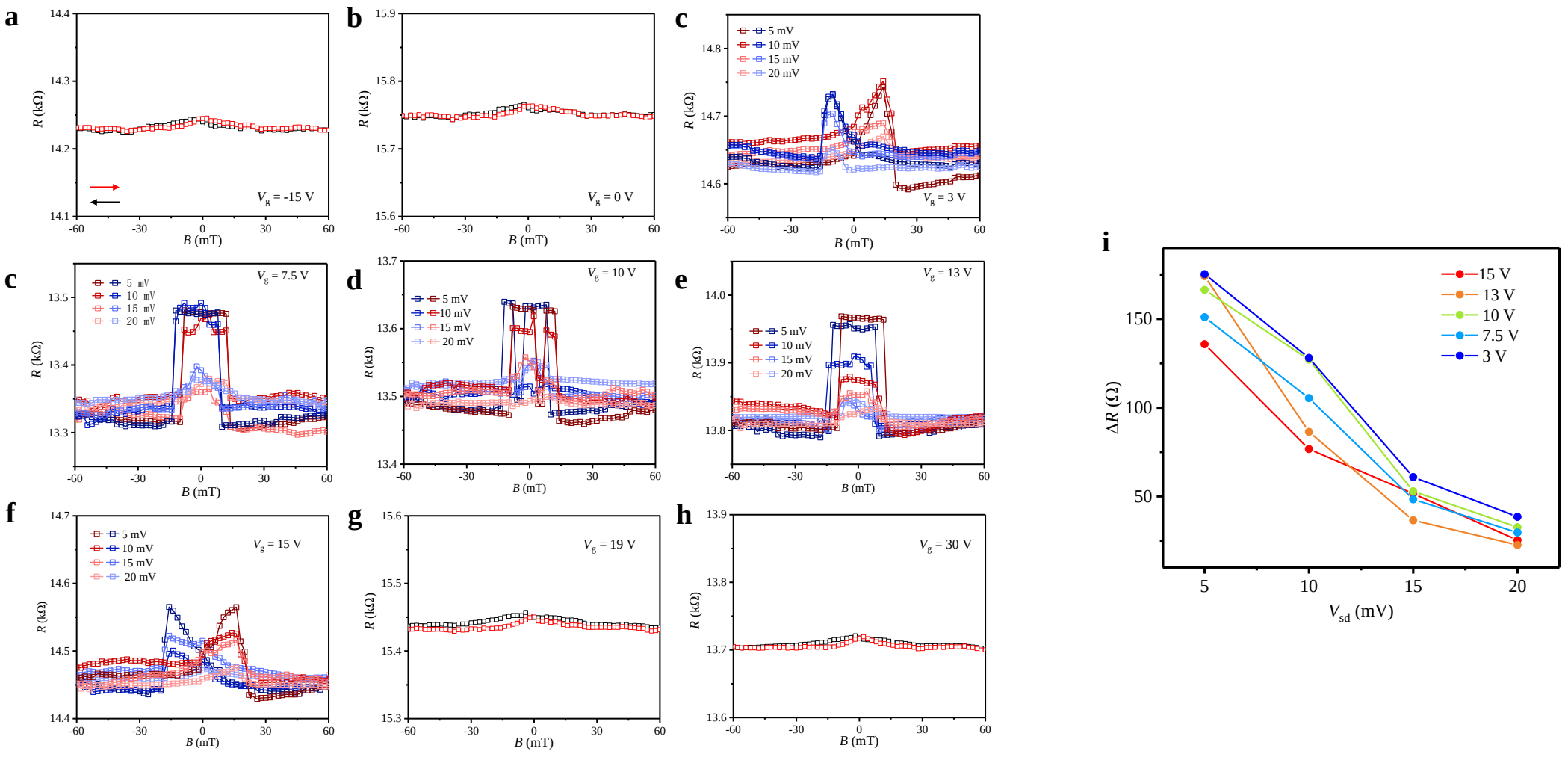

**Figure S9. Bias dependent magnetoresistance of the zGNR FET under different $V_{gate}$ at $T = 4$ K.** Magnetoresistance at $V_{sd} = 5$ mV, 10 mV, 15 mV and 20 mV under (a-h) different $V_{gate}$. The arrow represents the sweep direction of the magnetic field. (i) $V_{sd}$ dependence of magnetoresistance change ($\Delta R$), under different $V_{gate}$. $\Delta R$ is obtained by subtracting a polynomial fitting from the device resistance.

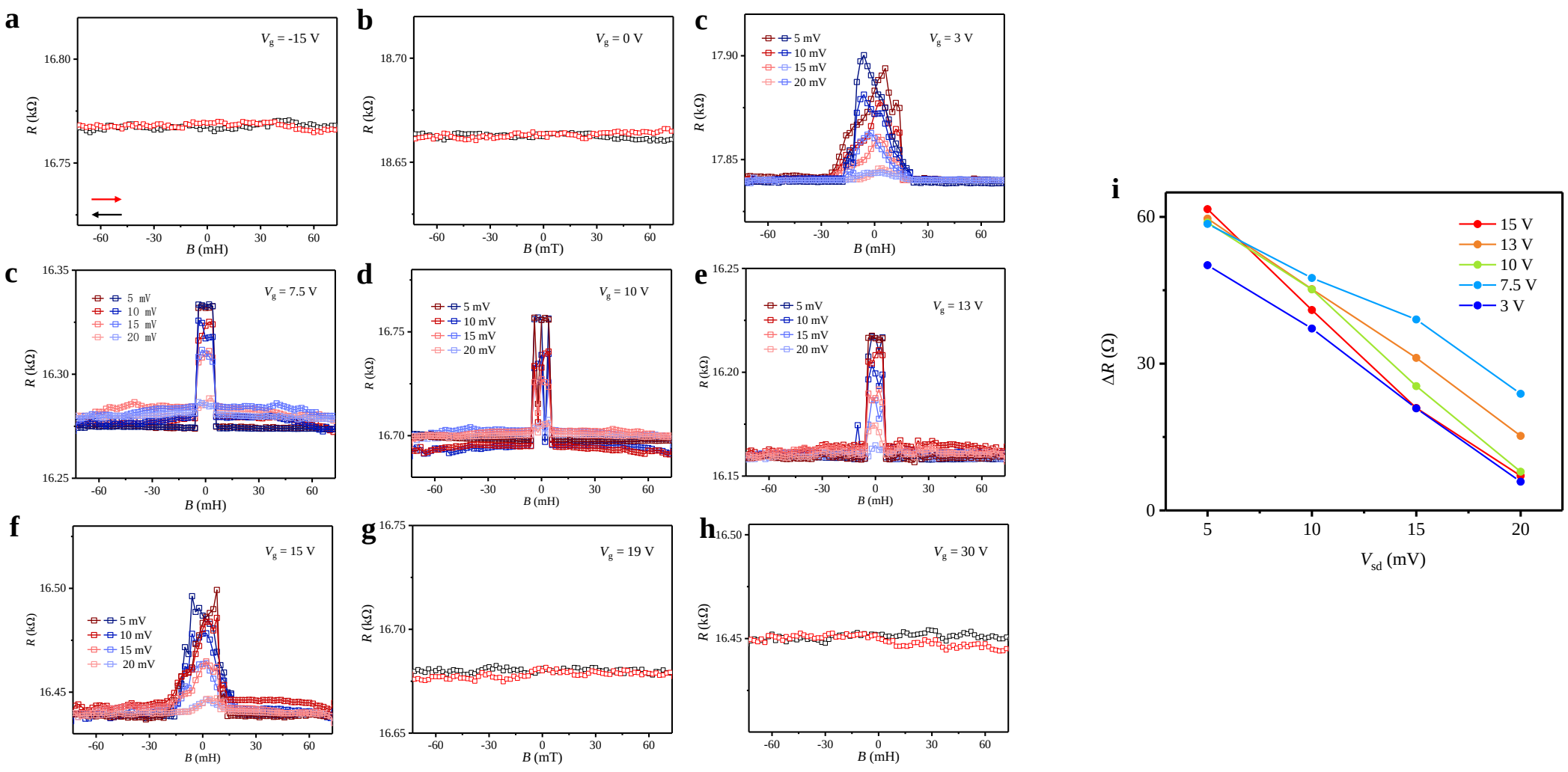

**Figure S10. Bias dependent magnetoresistance of the zGNR FET under different $V_{gate}$ at $T$ = 300 K.** Magnetoresistance at $V_{sd}$ = 5 mV, 10 mV, 15 mV and 20 mV under (a-h) different $V_{gate}$. The arrow represents the sweep direction of the magnetic field. (i) $V_{sd}$ dependence of magnetoresistance change ($\Delta R$). $\Delta R$ is calculated by subtracting a polynomial fitting from the device resistance.

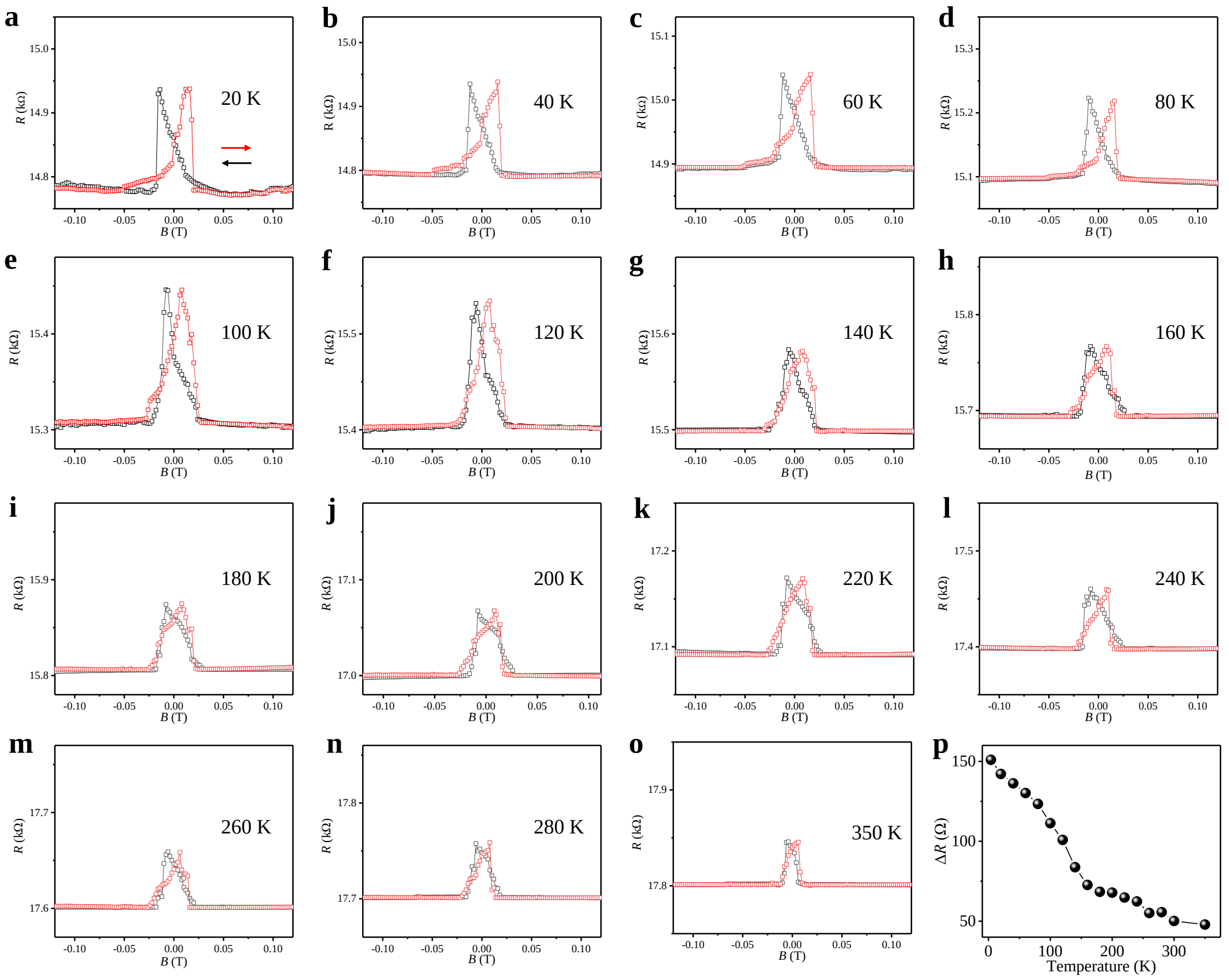

**Figure S11.** Temperature dependent magnetoresistance of the zGNR FET at different temperature (a-n) under $V_{\mathrm{gate}} = 3$ V and $V_{\mathrm{sd}} = 5$ mV. The red (black) arrow represents the sweep direction of the magnetic field. (o) Temperature dependent magnetoresistance change ($\Delta R$). $\Delta R$ is calculated by subtracting a polynomial fitting from the device resistance.

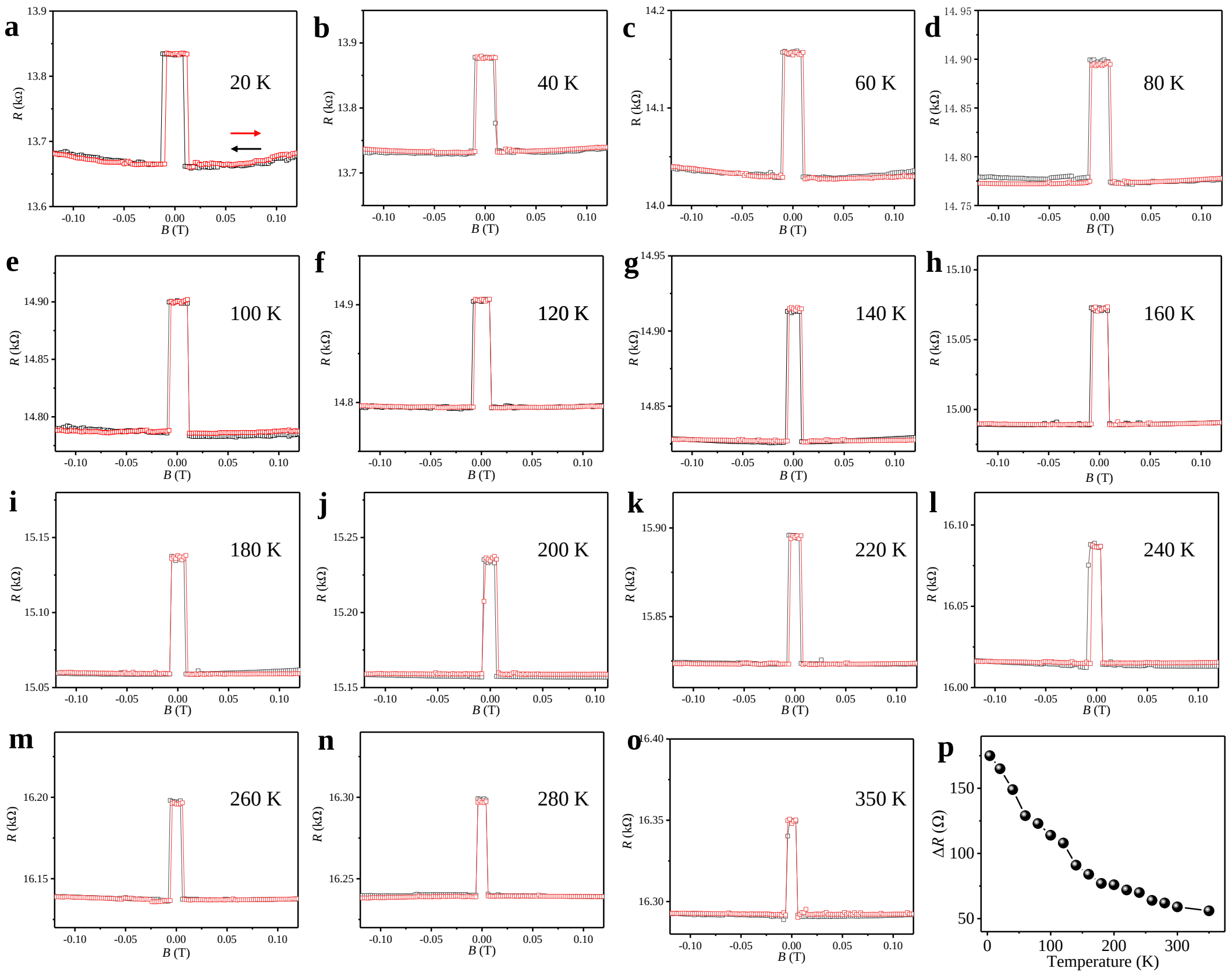

**Figure S12.** Temperature dependent magnetoresistance of the zGNR FET at different temperature (a-o) under $V_{gate} = 7.5$ V and $V_{sd} = 5$ mV. The red (black) arrow represents the sweep direction of the magnetic field. (p) Temperature dependent magnetoresistance change ($\Delta R$). $\Delta R$ is calculated by subtracting a polynomial fitting from the device resistance.